\documentclass[a4paper,reqno,12pt]{amsart}
\usepackage{amssymb,euscript,bbold}
\usepackage{array}
\usepackage{epsfig}
\renewcommand{\d}{\partial}
\newcommand{\half}{\tfrac12}
\newcommand{\reg}{\text{reg}}
\renewcommand{\gg}{{\ensuremath{\mathfrak g}}}

\newcommand{\sG}{{\ensuremath{\mathsf G}}}
\newcommand{\sT}{{\ensuremath{\mathsf T}}}
\newcommand{\sJ}{{\ensuremath{\mathsf J}}}

\newcommand{\1}{{\ensuremath{\mathbf 1}}}

\newcommand{\II}{\ensuremath{I\!I}}

\DeclareMathOperator{\ord}{\mathrm{ord}}
\begin{document}

\title[]{D-branes in curved spacetime:  Nappi--Witten background${}^*$} 
\author[]{Sonia Stanciu${}^1$ \ and \  Arkady  Tseytlin$^{1,2}$}
\address[]{\begin{flushright}${}^1$Theoretical Physics Group\\
Blackett Laboratory\\
Imperial College\\
London SW7 2BZ, UK\\
${}^2$Institute of Theoretical Physics\\
University of California\\
Santa Barbara, CA 93106, USA
\end{flushright}}
\email{s.stanciu@ic.ac.uk, tseytlin@ic.ac.uk}
\thanks{${}^*$ Imperial/TP/97-98/40, NSF-ITP-98-052, {\tt hep-th/9805006}}
\begin{abstract}
We find exact D-brane configurations in the Nappi--Witten background using
the boundary state approach and describe how they are related by T-duality
transformations.  We also show that the classical boundary conditions of the
associated sigma model correspond to a field dependent automorphism relating
the chiral currents and discuss the correspondence between the boundary state
approach and the sigma model approach.
\end{abstract}
\maketitle

\section{Introduction}

The {\it boundary state} formalism (see, e.g.,
\cite{CLNY,PC,Li,CK}) has become in the last years one of the
main approaches to the study of D-branes \cite{P} in various type II
string backgrounds \cite{OOY, BBMOOY, KO, SDKS, RS, FuSch}.  In
particular, this approach showed that D-branes probe a new aspect of
the background geometry, namely the geometry of submanifolds.

However, despite its success in unravelling some of the structure
underlying D-brane geometry, the geometric information obtained
through the boundary state approach is often rather difficult to
interpret in terms of the more standard sigma model approach.  For
instance, in the Calabi-Yau case the geometric picture of D-branes
wrapping around supersymmetric cycles emerges only in the large-volume
approximation in which the metric is taken to be essentially flat.  In
the case of group or coset spaces this connection seems even more
difficult to attain, as the fields in terms of which the conformal
structure is realised do not seem to have an obvious geometric (that
is, spacetime) interpretation.

In order to make progress in the understanding of D-branes in curved
spaces one needs to somehow bridge the gap between the boundary state
approach and the corresponding sigma model interpretation.  In this
paper, we attempt to do this, by choosing a simple string background
\cite{NW} and studying the corresponding D-brane configurations from
both points of view.

In order to apply the boundary state approach one has to start with an
exact string solution whose 2d CFT description is explicitly known.
On the other hand, in order to give D-branes a spacetime
interpretation we need to consider a string background which has a
sigma model realisation.  One of the few known classes of exact string
backgrounds (see, e.g., \cite{TT}) with spacetime interpretation is
represented by WZW models \cite{WW}.  In this case spacetime is a
group manifold and the exact (non-perturbative) conformal invariance
is guaranteed by the Sugawara construction.\footnote{As in the
previous studies, we ignore back reaction of D-branes on spacetime,
i.e. investigate possible D-branes that can be embedded in a fixed
geometry. In some cases (representing BPS superpositions of several
branes) one knows how the supergravity solution changes once one adds
a new D-brane. In (most of) these cases, however, one does not have a
2d CFT description of the corresponding string solutions and thus of
the corresponding D-branes.}

We have chosen the Nappi-Witten (NW) solution \cite{NW} for the two
reasons.  From the point of view of the boundary state approach, the
simplicity of the corresponding WZW model allows us to solve exactly
for the boundary conditions and hence find explicitly the D-brane
configurations.  On the other hand, despite its simplicity, this model
describes a curved homogeneous four-dimensional spacetime with
Minkowski signature.  This distinguishes our analysis from previous
studies in which D-branes in compact spaces were considered.  Here the
D-branes {\it worldvolumes} will be fully embedded in the NW
spacetime.  For the sake of simplicity we consider mainly the bosonic 
case (generalisation to supersymmetric case is straightforward).

The paper is organised as follows.  We start, in Section 2, by
reviewing the NW background and the underlying CFT.  In Section 3 we
describe the boundary state approach to finding D-brane configurations
in a closed-string background.  We write down the appropriate boundary
conditions and solve them explicitly, obtaining two classes of
solutions, each of them described by a matrix depending on three real
parameters.  Further, in Section 4, we study the geometry of these
solutions, and find that one of the two classes describes D3- and
D1-branes, whereas the other one describes euclidean D0-branes.  In
Section 5 we consider T-duality transformations in the space of
boundary states, and explain how they map between various D-branes.

In Section 6 we re-interpret these D-brane configurations as static
classical solutions of the Born-Infeld action for a D-brane probe in
the NW background.

We then turn, in Section 7, to the sigma model approach.  We start
with a detailed analysis of the boundary conditions of general 2d
sigma models and show that the standard boundary conditions in terms
of fields can be recast in terms of the chiral currents.  This is
particularly important for the WZW models, where the chiral currents
are conserved and generate, upon quantisation, the affine algebra
which underlies the conformal invariance of the model.  We then
specialise to NW background and conclude that these classical boundary
conditions obtained from the sigma model action are described by a
matrix which is closely related to the one obtained in Section 3 using
the boundary state approach, with the essential property that the
parameters are actually field-dependent functions.

Some details of the space-time embedding of the D-string solution are
given in Appendix A.  In Appendix B we briefly discuss the
supersymmetric generalisation of the construction of the boundary
states in the Nappi-Witten model.

\section{The Nappi--Witten background}

The Nappi-Witten background \cite{NW} (for generalisations of this
model see for instance \cite{ORS,KK,KKL}) is an exact four-dimensional
(super)string solution defined by the following WZW action
\begin{equation}
I[g] = \int_{\Sigma} \langle g^{-1}\d g, g^{-1}\bar\d
                g\rangle + {\textstyle \frac{1}{6} }\int_B \langle g^{-1}dg,
                [g^{-1}dg,g^{-1}dg]\rangle~,\label{eq:wzw} 
\end{equation}
where the fields $g$ are maps from a closed orientable Riemann surface
$\Sigma$ to the Lie group $\mathbf{G}$, which is to be thought of as the
simply-connected group corresponding to the $d=4$  Lie algebra $\gg$
with the generators $X_a=$ ($P_1$, $P_2$, $J$, $K$)  satisfying 
\begin{equation}\label{eq:gg}
[P_1, P_2] = K~,\qquad [J, P_1] = P_2~,\qquad [J, P_2] = -P_1~.
\end{equation}
This algebra is a central extension of the two-dimensional Poincar\'e
algebra. Its important feature is that it possesses an invariant
metric $\langle X_a,X_b\rangle = G_{ab}$ given by 
\begin{equation*}
G = \begin{pmatrix}
	 1 & 0 & 0 & 0\\
         0 & 1 & 0 & 0\\
         0 & 0 & b & 1\\
         0 & 0 & 1 & 0
	 \end{pmatrix}~,
\end{equation*}
where  $b$ is a  real parameter (which can be, in principle, absorbed in
 a redefinition of the generators).

The group manifold $\mathbf{G}$ can be parametrised as follows
\begin{equation}
g = e^{a_i P_i}e^{uJ + vK}~,\label{eq:parg}
\end{equation}
where $(a_1,a_2,u,v)$ play the r\^ole of the spacetime coordinates in 
 string theory.  In terms of these fields (1) becomes  a sigma-model 
 action, with the spacetime metric  and 2-form given by 
\begin{equation}\label{eq:met}
ds^2 = da_i da_i  - \epsilon_{ij} a_i da_j du + b du^2+ 2du dv ~,
\end{equation}
\begin{equation} 
B = \half\epsilon_{ij}a_i da_j du~.\label{eq:B}
\end{equation}
This background describes a gravitational plane wave in $d=4$
space-time with signature $(+++-)$. 

The exact conformal invariance of this model is based, as is well known, on
its infinite-dimensional symmetry group $\mathbf{G}(z)\times\mathbf{G}(\bar
z)$ characterised by the conserved currents
\begin{equation}
J(z)=-\d g g^{-1}~,\qquad \bar J(\bar z)=g^{-1}\bar\d g~.\label{eq:cc}
\end{equation}
These currents generate an affine Lie algebra $\widehat\gg$ described
by  
\begin{equation}
J_a (z)J_b (w) = \frac{G_{ab}}{(z-w)^2} + \frac{{f_{ab}}^c J_c
(w)}{z-w} + \reg~,\label{eq:affg}
\end{equation}
in the holomorphic sector, and similar OPEs in the antiholomorphic
sector.  Furthermore, the generalisation of the Sugawara construction
to the nonsemisimple algebra case \cite{NW,Msug,FSsug,FS2} gives us a
CFT with the central charge equal to four and the energy-momentum
tensor 
\begin{equation*}
\sT(z) = \Omega^{ab}(J_aJ_b)(z)~,
\end{equation*}
where $\Omega^{ab}$ is the inverse of the invariant metric,
$\Omega_{ab} = 2G_{ab} + \kappa_{ab}$, and $\kappa$ is the
Killing form of $\gg$:\footnote{It is not a priori clear that
$G$ and $\Omega$ can simultaneously be nondegenerate.  However, 
this is true for self-dual Lie algebras \cite{FS3}, and our $\gg$
falls in this class.}
\begin{equation*}
\Omega = \begin{pmatrix}
	 2 & 0 &   0  & 0\\
         0 & 2 &   0  & 0\\
         0 & 0 & 2b-2 & 2\\
         0 & 0 &   2  & 0
	 \end{pmatrix}~.
\end{equation*}

\section{Boundary states}

The approach of constructing boundary states of closed string theories
(see, e.g., \cite{CLNY}) is based on the requirement of conformal
invariance.  In open string theories one has to impose constraints on
the boundary conditions such that the conformal symmetry is not
broken.  Then the boundary can be thought of as a closed string state
where the left-- and right--moving conformal structures are related in
a {\em consistent\/} way.  The strategy is basically to try to find
the consistent boundary conditions on the fields, preserving the
``overall'' conformal structure.

In the case of a background described by a WZW model \cite{I,KO,SDKS}
the natural variables, that is, the fields in terms of which the
conformal structure of the model is realised,  are the affine currents.
Hence it is on them that one imposes the  boundary conditions 
\begin{equation}
J_a(z) + {R^b}_a {\bar J}_b(\bar z) = 0~.\label{eq:bc} 
\end{equation}
These boundary conditions have
to satisfy the following  consistency requirements:
\begin{itemize}
\item[(i)] They  preserve conformal invariance, that is
\begin{equation*}
\sT(z) - \bar\sT(\bar z) = 0~,\label{eq:ci}
\end{equation*}
at the boundary.\footnote{More generally, the holomorphic and the
antiholomorphic sectors are related by an automorphism of the
corresponding CFT.  However,  since the automorphism group of the
Virasoro algebra is trivial,  we have the above condition.}
\item[(ii)] They preserve the infinite-dimensional symmetry of the
current algebra \eqref{eq:affg}.  One can alternatively argue that
this condition is imposed by the fact that the boundary condition for
the energy-momentum tensor is not restrictive enough to determine the
allowed configurations uniquely.
\end{itemize}

In order to analyse these conditions, it is convenient to define the
map $R:\gg\to\gg$, defined as $R(X_a)=X_b{R^b}_a$, where $X_a\in\{P_1,
P_2, J, K\}$.  Then the first requirement translates into 
\begin{equation}
R^T \Omega R = \Omega~,\label{eq:rmet}
\end{equation}
whereas the second one imposes on $R$ the 
two conditions, corresponding to
the first and second order pole of the OPE in \eqref{eq:affg},
respectively:  
\begin{equation}
[R(X_a),R(X_b)] = R([X_a,X_b])~,\qquad
R^T G R = G~.\label{eq:condr}
\end{equation}
Notice that \eqref{eq:rmet} and the second condition in
\eqref{eq:condr} are equivalent in the particular case of our Lie
algebra $\gg$, because of the special form of the metric and of the
Killing form.  We are therefore left with only two conditions, stating
that $R$ must be an {\it automorphism} of $\gg$ which {\it preserves
the metric}.  

One of the appeals of the NW  model is that it is simple enough to allow
us to solve for these conditions explicitly.  We obtain two families
of solutions: the first 
\begin{equation}\label{eq:RI}
R_I = \begin{pmatrix}
	 \cos\phi & -\sin\phi & r_1 & 0\\
         \sin\phi &  \cos\phi & r_2 & 0\\
           0    &    0    &  1  & 0\\
         -r_1\cos\phi -r_2\sin\phi & r_1\sin\phi -r_2\cos\phi &
                                   -\frac{1}{2}(r_1^2 + r_2^2) & 1
	 \end{pmatrix}
\end{equation}
and  the second 
\begin{equation}\label{eq:RII}
R_{\II} = \begin{pmatrix}
	 \sin\phi &  \cos\phi & r_1 & 0\\
         \cos\phi & -\sin\phi & r_2 & 0\\
             0    &      0    &  -1 & 0\\
         r_1\sin\phi +r_2\cos\phi & r_1\cos\phi -r_2\sin\phi & 
	 \frac{1}{2}(r_1^2 + r_2^2) & -1
	  \end{pmatrix}
\end{equation}
Both   solutions depend on
three real parameters, $r_1$,$r_2$ and $\phi$. In fact,  the ``moduli
space'' in each case is  $\mathbb{R}^2\times S^1$.  As follows from
(8), $(\det R)^2=1$.  Indeed, 
we find  that $\det R_I=1$,\  $\det R_{\II}  =-1$. This suggests that
$R_I$ ($R_{\II}$) will describe odd (even) dimensional D-branes. 

One can  represent $R_I$ in a more compact form as
\begin{equation*}
R_I = Ad (e^{\phi J} e^{\epsilon_{ij}r_i P_j})~,
\end{equation*}
which shows that $R_I$ is given by the inner automorphisms of $\gg$.
$R_{II}$  can be related to $R_I$ by noticing that
\begin{equation*}
R_I(\phi,r_1,r_2) = \begin{pmatrix}
                          0 & 1 & 0 & 0\\
                          1 & 0 & 0 & 0\\
                          0 & 0 & -1 & 0\\
                          0 & 0 & 0 & -1
	            \end{pmatrix} R_{\II}(\phi,r_2,r_1)~.
\end{equation*}

\section{D-brane solutions and geometry}

So far we have determined the boundary states of the Nappi--Witten
model which preserve conformal invariance and the algebraic structure
underlying it.  However,  it is not clear that any such boundary state
will have a {\it geometric interpretation} as a D-brane.  First,  we
have to clarify the way in which the geometric information defining
the D-brane worldvolume is encoded in our boundary conditions.  In
other words,  we have to define what we mean by Neumann and Dirichlet
boundary conditions.  This step is particularly important in our case,
since the boundary conditions are not imposed directly on the fields
but rather on the {\it chiral currents}. 

The best way of deciding which boundary condition is a Neumann one and
which is a Dirichlet one is to evaluate $J$ and $\bar J$ at the origin
(or in the flat space limit).  We find that at $a_1 = a_2 = u = v = 0$
\begin{align*}
J|_0 &= - \d a_1 P_1 - \d a_2 P_2 - \d u J - \d v  K ~,\\
\bar J|_0 &= \bar\d a_1 P_1 + \bar\d a_2 P_2 + \bar\d u J + \bar\d v
K~.
\end{align*}
This shows that, for the currents, the Neumann boundary conditions for
the fields correspond to $J_a = -\bar J_a$, whereas
the Dirichlet one -- to $J_a = \bar J_a$.

To analyse the boundary states we have obtained and identify the
D-branes they describe we have to study the eigenvalues and
eigenvectors of the linear operator $R$.  We shall consider the two
classes of solutions separately.

\subsection{$R_I$ boundary states}

The eigenvalues of $R_I$ are given by 
\begin{equation*}
\lambda_{1,2} = e^{\pm i\phi}~, \qquad \lambda_3 = \lambda_4 = 1~.
\end{equation*}
In fact, $R_I$ can be brought, by a real change of basis, to the
following ``standard'' form 
\begin{equation*}
R_I^{st} = \begin{pmatrix}
	 \cos\phi & -\sin\phi & 0 & 0\\
         \sin\phi &  \cos\phi & 0 & 0\\
             0    &     0     & 1 & 0\\
             0    &     0     & 0 & 1
	   \end{pmatrix}~, 
\end{equation*}
which makes it easy to identify the D-brane configurations $R_I$
describes. 

For $\phi\neq \pi$ our solution describes a {\it D3-brane} with
``generalised'' Neumann (or ``mixed") boundary conditions along the
first two directions (for $\phi=0$, $R_I$ becomes simply the identity
matrix, which is equivalent to having Neumann boundary conditions in
all 4 directions).  One may be tempted to conclude that we have a
background gauge field at the boundary (cf. \cite{P,BDL}) whose field
strength depends on $\phi$.  However, since we are dealing with a WZW
model, a non-trivial 2-form field $B_{ab}$ may account for the above
boundary conditions.  We shall come back to this point in Section 6
where we will analyse the boundary conditions from the point of view
of the corresponding sigma model.

In the case of $\phi=\pi$ the eigenvalues of $R_I$ read
\begin{equation*}
\lambda_1 = \lambda_2 = -1~,\qquad \lambda_3 = \lambda_4 = 1~,
\end{equation*}
so that this boundary state describes a {\it D1-brane} (D-string).  The
tangent space of its worldsheet is spanned by the eigenvectors of $R_I$
corresponding to the $+1$ eigenvalues:
\begin{equation*}
Y_3 = \half r_1P_1 + \half r_2P_2 + J~, \qquad 
Y_4 = K~.
\end{equation*}
On the other hand, the eigenvectors corresponding to the Dirichlet
directions read 
\begin{equation*}
Y_1 = P_1 - \half r_1K~, \qquad 
Y_2 = P_2 - \half r_2K~.
\end{equation*}
One can now check that the tangent vectors to the Neumann directions
form an abelian subalgebra, which we will write formally as
$[N,N]\subset N$.  This means that the corresponding worldsheet is
actually a submanifold of the target manifold.  Moreover the tangent
vectors to the Dirichlet directions satisfy $[D,D]\subset N$, as
expected.

Notice that in this case $R_I^2 = \1$, which implies that the metric
splits with respect to the directions tangent and normal to the
worldsheet of the D-string
\begin{equation*}
\Omega = \Omega_N + \Omega_D~,
\end{equation*}
where $\Omega_N$ denotes the induced metric on the tangent space to
the D-brane and $\Omega_D$ is the metric induced in the normal bundle.

We can go further and obtain a spacetime description of this D-string
solution.  The generic tangent vector to the D-string worldsheet is of
the form $\alpha Y_3 +\beta Y_4$, with $\alpha$ and $\beta$ arbitrary
parameters.  This generates a surface which is nothing but the
worldsheet of the D-string.  On the other hand, every point on this
surface can be parametrised as a group element
\eqref{eq:parg}. This  yields an equation
\begin{equation*}
e^{\alpha Y_3 + \beta Y_4} = e^{a_i(\alpha,\beta)P_i}
e^{u(\alpha,\beta)J + v(\alpha,\beta)K}~,
\end{equation*}
determining  the spacetime fields $a_i,u,v$ in terms of the parameters 
$\alpha,\beta$, i.e.  
 the surface describing the D-string worldsheet, 
\begin{align} 
a_1(\alpha,\beta) &= \half r_1\sin\alpha + \half r_2
                     (\cos\alpha-1)~,\label{eq:D1a}\\
a_2(\alpha,\beta) &= -  \half r_1  (\cos\alpha-1) + \half r_2
                     \sin\alpha~,\label{eq:D1b}\\
u(\alpha,\beta) &= \alpha~,\label{eq:D1c}\\
v(\alpha,\beta) &= \beta + {\textstyle {\frac{1}{8}}}(r_1^2+r_2^2)
                   (\alpha-\sin\alpha)~. \label{eq:D1d} 
\end{align}
Using the expression for the spacetime metric \eqref{eq:met} 
one
can now compute the induced metric on the D-string world sheet:
\begin{equation}\label{eq:Dmet} 
ds^2 = \left[b + \tfrac14 (r_1^2+r_2^2)\right]d\alpha^2 +
       2d\alpha d\beta~. 
\end{equation}
Notice that in the particular case of  $r_1=r_2=0$ we have 
\begin{equation}
a_1(\alpha,\beta) = a_2(\alpha,\beta) = 0~,\quad 
u(\alpha,\beta) = \alpha~,\quad
v(\alpha,\beta) = \beta~,\label{eq:D1}
\end{equation}
which means that the worldsheet of the corresponding D-string
coincides with the $(u,v)$ plane.

The metric \eqref{eq:Dmet} describes a two-dimensional Minkowski spacetime,
and therefore the worldsheet of the D-string is flat in the induced metric.
Moreover it is easy to see that its extrinsic curvature also vanishes. This
is because the worldsheet is actually the manifold of a Lie subgroup
$\mathbf{H}$ of $\mathbf{G}$ and relative to a bi-invariant metric on
$\mathbf{G}$, any Lie subgroup $\mathbf{H}$ is totally geodesic.  This is
equivalent to the vanishing of the second fundamental form and therefore of
the extrinsic curvature.

\subsection{$R_{\II}$ boundary states}

$R_{\II}$ case can be analysed along similar lines.  The eigenvalues 
turn out to be 
completely independent of the parameters, 
\begin{equation*}
\lambda_1 = 1~, \qquad \lambda_2 = \lambda_3 = \lambda_4 = -1~.
\end{equation*}
The Neumann eigenvector always exists and is given by
\begin{equation}\label{eq:y1}
Y_1 = \begin{cases}
      (1+\sin\phi) P_1 + \cos\phi P_2\\
      \phantom{asdasd} + \half [r_1(1+\sin\phi) +
       r_2\cos\phi] K & \text{for $\phi\neq \tfrac{3\pi}2$;}\\
      P_2 + \half r_2 K & \text{otherwise.}
      \end{cases}
\end{equation}
As can be easily checked, it has positive norm.  This means that at
any point on the integral curve of this vector field, the tangent
space to the spacetime has an orthogonal decomposition into a
1-dimensional subspace tangent to the curve and a 3-dimensional
subspace normal to it.  The subspace tangent to the curve is the
eigenspace of $R_{\II}$ with eigenvalue $+1$. However, a generic
$R_{\II}$ will not act like $-\1$ on the subspace normal to the curve,
i.e., it will not give rise to the Dirichlet boundary conditions.  For
this to be the case we will be forced to restrict the values of $r_1$,
$r_2$, and $\phi$.

Indeed, $R_{\II}$ has always one Dirichlet eigenvector $Y_2$ given by
$K$.  The existence of the other two Dirichlet eigenvectors requires
the diagonalisability of $R_{\II}$.  This in turn requires that the
parameters obey the following relation:
\begin{equation}\label{eq:diagcond}
r_1\left(\cos{\tfrac\phi2} - \sin{\tfrac\phi2}\right) = 
r_2\left(\cos{\tfrac\phi2} + \sin{\tfrac\phi2}\right)~.
\end{equation}
In this case we can conclude that these boundary states describe a
{\it D0-brane} whose tangent space is spanned by $Y_1$.  Since $Y_1$
has positive norm, we deduce that it is a euclidean  D0-brane.  We can
distinguish four cases; in each of these cases one can write  down the
corresponding eigenvectors.  $Y_1$ is given by \eqref{eq:y1} and
$Y_2=K$.  The other two Dirichlet eigenvectors $Y_3$ and $Y_4$ are
given in the following table.

\begin{table}[h!]
\renewcommand{\arraystretch}{1.5}
\begin{tabular}{|>{$}c<{$}|>{$}c<{$}|>{$}c<{$}|}
\hline
(\phi,r_1,r_2) & Y_3 & Y_4\\
\hline
\phi\neq \tfrac\pi2,\tfrac{3\pi}2,~r_1r_2\neq 0 &
P_1 - \frac{2r_1}{r_1^2 + r_2^2} J &
P_2 - \frac{2r_2}{r_1^2 + r_2^2} J\\
\phi\neq \tfrac\pi2,\tfrac{3\pi}2,~r_1 = r_2 = 0 &
P_1 - \frac{\cos\phi}{1 - \sin\phi}J & J\\
\phi=\tfrac\pi2,~r_2=0 & P_2 & J - \half r_1 P_1\\
\phi=\tfrac{3\pi}2,~r_1=0 & P_1 & J - \half r_2 P_2\\
\hline
\end{tabular}
\end{table}

In this case one can also check that $[N,N]\subset N$ (trivially,
since $N$ is one-dimensional) and that $[D,D]\subset N$.  Moreover, in
each of the cases listed above, $R_{\II}^2=\1$ and the metric splits
as
\begin{equation*}
\Omega = \Omega_N + \Omega_D~.
\end{equation*}
The spacetime description of this D0-brane configuration can be
determined by parametrising the world-line of the brane by $\alpha$,
such that
\begin{equation*}
e^{\alpha Y_1} = e^{a_i(\alpha)P_i} e^{u(\alpha)J + v(\alpha)K}~,
\end{equation*}
which allows us to write down the explicit expressions for the curve.
For $\phi\neq\tfrac{3\pi}2$ we find
\begin{align}
a_1(\alpha) &= (1+\sin\phi)\alpha\label{eq:D0a}\\
a_2(\alpha) &= \cos\phi~ \alpha\label{eq:D0b}\\
u(\alpha) &= 0\label{eq:D0c}\\
v(\alpha) &= \half\left[r_1(1+\sin\phi) + r_2
\cos\phi\right]\alpha~.\label{eq:D0d} 
\end{align}
The induced metric on the world line  is given by $ds^2 =
2(1+\sin\phi)d\alpha^2~.$  For $\phi=\tfrac{3\pi}2$ the world line of
the D0-brane is described by  
\begin{equation}\label{eq:D0}
a_1(\alpha) = 0~,\quad 
a_2(\alpha) = \alpha~,\quad 
u(\alpha) = 0~,\quad
v(\alpha) = \half r_2 \alpha~.
\end{equation}
and the corresponding expression for its induced metric is $ds^2 =
d\alpha^2~.$

\section{ T-duality}

In this section we shall study the effect of abelian T-duality
transformations on the D-brane configurations we have found above.
The structure of the duality group here is slightly different from the
one in the case of flat backgrounds.  In particular, the WZW model is
{\em self-dual\/} under an abelian T-duality transformation
\cite{K,AAL,GW}. 

By a T-duality transformation we will understand a map \cite{KO,SDKS},
at the level of the fields, which preserves the conformal structure of
the model.  More precisely, we will consider a map that acts trivially
on the antiholomorphic sector of the theory and preserves the Virasoro
algebra of the holomorphic sector.\footnote{In principle, one should
demand that T-duality acts as an automorphism on the CFT of the
holomorphic sector, but the automorphism group of the Virasoro algebra
is trivial.}  Hence if the original theory is described by the
currents $J_a(z)$, $\bar J_a(\bar z)$, the dual theory will be
described by
\begin{equation*}
J'_a = {T^b}_a J_b~,\qquad {\bar J}'_a = {\bar J}_a~,
\end{equation*}
where the map $T: \gg\to\gg$ is defined by $T(X_a)={T^b}_a X_b$. 

An arbitrary T-duality transformation is defined by the properties:
\begin{itemize}
\item[(i)] It preserves the conformal structure of the model.  Since
the antiholomorphic CFT will be trivially preserved, this means that
the T-duality map at the level of the currents will have to preserve
the Sugawara  energy-momentum tensor of the holomorphic sector 
\begin{equation*}
\sT(z) = \sT'(z)~.
\end{equation*}
\item[(ii)] It preserves the infinite-dimensional symmetry of the
current algebra \eqref{eq:affg}. 
\end{itemize}

This group of transformations is characterised by a set of generators
which satisfy, in addition,
\begin{itemize}
\item[(iii)] $T^2 = \1~.$
\end{itemize}
Then any T-duality transformation can be written as a product of a
certain number of generators.

The analysis of the linear map $T$ is very similar, at a formal level,
to the one of the map $R$ (see Section 3) which describes the boundary
conditions: the first two requirements will make the matrix $T$ take
exactly the same form as the matrix $R$.  If we further impose that
$T^2 = \1$ we obtain the two families of (non-trivial) solutions.  The
first one is given by
\begin{equation*}
T_I = \begin{pmatrix}
	 -1 &  0 & t_1 & 0\\
          0 & -1 & t_2 & 0\\
          0 &  0 &  1  & 0\\
         t_1& t_2& -\frac{1}{2}(t_1^2 + t_2^2) & 1
	 \end{pmatrix}~.
\end{equation*}
It is a two-parameter family of T-duality transformations.  If we
define the order of $T$ to be equal to the number of $-1$ eigenvalues
then  $\ord(T_I) = 2$.  The second class of solutions is given by
\begin{equation*}
T_{\II} = \begin{pmatrix}
	 \sin\theta &\cos\theta &  t_1 & 0\\
         \cos\theta & -\sin\theta &  t_2 & 0\\
           0    &    0    &   -1 & 0\\
          t_1   &  t_2    &  \half(t_1^2 + t_2^2) & -1
	 \end{pmatrix}~,
\end{equation*}
and hence  $\ord(T_{\II}) = 3$.

In general, if we start with a boundary state characterised by the
matrix $R$ and perform a T-duality transformation given by the
matrix $T$, the T-dual configuration will be described by the matrix
\begin{equation*}
R' = R \ T~.
\end{equation*}

Because both $R$ and $T$ are in the (double cover of the) adjoint
group, so is $R'$, and, moreover, given $R'$ and $R$ there exists a
$T$ which relates them, namely $R^{-1}R'$.  Therefore, any two
boundary states will be related by a T-duality.  For example, the
D-string we analysed before, which is defined by $R_I(\pi,r_1,r_2)$,
can be described as the T-dual of the D3-brane defined by $R_I=\1$,
where the T-duality transformation is given by
$T=R_I(\pi,-r_1,-r_2)$.  Similarly, a D0-brane configuration, say
$R_{II}(\tfrac{\pi}{2},0,0)$, can be obtained, via a transformation
$T_{\II}=R_{II}$, from the D3-brane configuration $R_I=\1$. 

We now consider the two types of duality transformations separately.

\subsection{$T_I$ duality transformations}

$T_I$ maps between boundary
states of the same type (described by matrices $R$ belonging to
the same class).  Indeed, if we start with a boundary state
characterised by the matrix $R_I(\phi, r_1, r_2)$ and apply the 
T-duality transformation characterised by  $T_I(\theta, t_1,
t_2)$ we obtain a (T-dual) boundary state  described by $R_I(\phi',
r'_1, r'_2)$ with
\begin{align*}
\phi' &= \phi + \theta~,\\
r'_1  &= r_1 + t_1 \cos\phi - t_2 \sin\phi~,\\
r'_2  &= r_2 + t_1 \sin\phi + t_2 \cos\phi~.
\end{align*}
In particular, we recover that the D-string and the D3-brane without a
background field are T-dual to each other.

On the other hand, if we start with a boundary state described by
$R_{\II}(\phi,r_1,r_2)$ and  perform the  transformation defined by
$T_I(\theta, t_1, t_2),$ we obtain a configuration described by 
$R_{\II}(\phi',r'_1,r'_2)$ with
\begin{align*}
\phi' &= \phi + \theta~,\\
r'_1  &= r_1 + t_1 \sin\phi + t_2 \cos\phi~,\\
r'_2  &= r_2 + t_1 \cos\phi - t_2 \sin\phi~.
\end{align*}

Here we have to make a few remarks.  We have seen in Section 4 that
$R_{\II}(\phi,r_1,r_2)$ is not diagonalisable in general, rather one
has to restrict the parameters to satisfy \eqref{eq:diagcond}.  It
turns out that generic $T_I(\theta,t_1,t_2)$ do not preserve this
condition.  If, on the other hand, we require that the dual
$R_{\II}(\phi',r'_1,r'_2)$ be diagonalisable as well, then we have to
restrict ourselves to the transformations $T_I(\theta,t_1,t_2)$ such
that $\phi'$, $r'_1$ and $r'_2$ satisfy a condition similar to
\eqref{eq:diagcond}.  It turns out that the only $T_I(\theta,t_1,t_2)$
for which this equation is satisfied for all $(\phi,r_1,r_2)$ obeying
\eqref{eq:diagcond} is the identity.  Hence there is no subgroup of
the T-duality subgroup which stabilises those boundary states with a
D-brane interpretation.

\subsection{$T_{\II}$ duality transformations}

Again we start by noticing that $T_{\II}$ maps between boundary states
of different type (described by $R$ matrices belonging to different
classes).  Thus if we consider a boundary state given by $R_I(\phi,
r_1, r_2)$ and apply a T-duality transformation characterised by the
matrix $T_{\II}(\theta,t_1,t_2)$ we obtain a T-dual boundary state
described by $R_{\II}(\phi', r'_1, r'_2)$ with
\begin{align*}
\phi' &= \theta - \phi\\
r'_1  &= -r_1 + t_1 \cos\phi - t_2 \sin\phi\\
r'_2  &= -r_2 + t_1 \sin\phi + t_2 \cos\phi~.
\end{align*}
Notice that the dual $R_{\II}$ will not, in  general,  be diagonalisable.
Thus  if we consider a D-brane configuration
$R_I$ and  perform a generic $T_{\II}$ transformation we will end up
with a boundary state described by a nondiagonalisable $R_{\II}$.  In
order for $R_{\II}(\phi',r'_1,r'_2)$ to be diagonalisable
$T_{\II}(\theta,t_1,t_2)$ will have to satisfy an extra condition
which follows from \eqref{eq:diagcond}.

Finally, let us  consider the effect of a $T_{\II}(\theta,t_1,t_2)$ on
a boundary state described by $R_{\II}(\phi,r_1,r_2)$.  The T-dual
configuration is given by $R_I(\phi',r'_1,r'_2)$ with 
\begin{align*}
\phi' &= \theta - \phi\\
r'_1  &= -r_1 + t_1 \sin\phi + t_2 \cos\phi\\
r'_2  &= -r_2 + t_1 \cos\phi - t_2 \sin\phi~.
\end{align*}

\section{D-brane configurations as classical solutions
of Born-Infeld action in NW background}

In Sections 3 and 4 we have discussed the requirements that a boundary
state has to satisfy in order to preserve (some of) the conformal
structure of the 2d theory and classified boundary states which can be
interpreted as D-brane configurations.  There is another approach to
determining which D-branes can be embedded in a given curved
background.  Ignoring back reaction, one may start with the standard
Born-Infeld action \cite{L} for a D$p$-brane probe moving in a curved
space\footnote{The NW background (times a torus) is an exact solution
of both bosonic or type II string theory, so the discussion that
follows applies to any of the two closed string theories.}  and find
if there are classical solutions describing static branes.  The
existence of a static solution (i.e.  the absence of a static
potential in the D-brane action) suggests that the corresponding brane
is a BPS state.\footnote{The BPS condition is usually phrased in terms
of the residual {\em spacetime supersymmetry\/} and is thus translated
(in the previously considered case of D-branes in the compactification
manifold) into boundary conditions imposed on the spectral flow
operator of the corresponding $N{=}2$ supersymmetric 2d theory
\cite{OOY,SDKS}. Residual supersymmetry condition \cite{BBS} implies
also satisfaction of the corresponding brane equations of motion.  The
`no-force' condition is, in principle, more general than that of the
residual supersymmetry, as it can be defined already in bosonic theory
(the condition of the absence of the static potential was used, e.g.,
in \cite{TBPS} to determine possible composite BPS configurations of
branes).}

Since the only closed string fields that are non-trivial in the NW
background are the metric and (NS-NS) antisymmetric tensor field, the
relevant part of the D-brane action is (we ignore higher-derivative
corrections) 
\begin{equation}\label{eq:DBI} 
I_p = \int d^{p+1} y~\sqrt{-\det(\hat G_{mn} +\hat B_{mn}+ F_{mn})}~, 
\end{equation}
where $x^\mu$ are the coordinates of the $D=10$ space and $y^m$ are
coordinates on the D$p$-brane world-volume and ($m,n=0,...,p$) 
\begin{equation*}
\hat G_{mn} = G_{\mu\nu}\d_m x^{\mu}\d_n x^{\nu},\quad 
\hat B_{mn} = B_{\mu\nu}\d_m x^{\mu}\d_n x^{\nu},\quad 
F_{mn}=\d_m A_n -\d_n A_m.
\end{equation*}
It is straightforward to check that in the cases of $p=3, \ p=1$ and $p=0$
the configurations $x^m=y^m,\ x^{p+1},...,x^{9}=0$, \ $F_{mn}=0$ are, indeed,
the solutions of the equations for $x^\mu$ and $A_m$ which follow from
(\ref{eq:DBI}).\footnote{The expansion of the action near the static brane
configuration starts with term linear in velocity, reflecting the `magnetic
field' interpretation of the off-diagonal component of $G+B$ matrix.}

In the case of the D3-brane probe action in the static gauge
($x^m=y^m$)  the equation for $A_m$ is equivalent to the `boundary'
condition of conformal invariance  of the corresponding string sigma
model (see below).

\section{Relation to the sigma model approach}

The boundary state approach can be applied to a wide range of string
backgrounds \cite{OOY, KO, SDKS, RS}, which generically need not have
a spacetime realisation in terms of sigma models.  Indeed, all the
necessary data is determined by the conformal structure of a
particular background.  Even in the cases in which an exact string
background possesses a sigma model realisation, the boundary state
configurations may not have sigma model description.  However, in
order to interpret the boundary state solutions in terms of D-brane
configurations in the target manifold one needs to understand the
geometric content of these states.  One should also keep in mind that
the method of boundary states has its limitations, as it is not
applicable to many string solutions described by conformal sigma
models for which one does not know explicitly the exact generators of
the (super)conformal algebra.

One of the remarkable features of the WZW models is that they fulfill
both requirements (explicitly known conformal structure and sigma
model realisation).  However, since the boundary conditions in the
boundary state approach are defined on the chiral currents rather than
on the fields there is an obvious lack of geometric interpretation of
the WZW boundary states, and in particular of the corresponding
D-brane configurations \cite{I, RS}.  Moreover, the Ansatz for the
relation between the currents at the boundary that one usually adopts
is purely linear, while one might think that this may not be a natural
assumption from the sigma model point of view, given that the
background is curved.

In this section we will analyse this problem.  We will start with a
general WZW model (although some of the expressions below will hold
for an arbitrary sigma model with 2-form field), and then we will
consider the particular case of the Nappi-Witten background.

Let us start with an action of a generic WZW model on a 2-space with a
disc topology with an additional interaction (1-form field $A$) at the
boundary
\begin{equation}
S = \int_{\Sigma} \langle g^{-1}\d g, g^{-1}\bar\d g\rangle + 
    \int_{\Sigma} g^* B + \int_{\d\Sigma} g^* A~.\label{eq:owzw}  
\end{equation}
Here the worldsheet $\Sigma$ is a two-dimensional manifold with
boundary $\d\Sigma$ and $B$ represents a particular choice for the
antisymmetric tensor field (see \cite{KlS})\footnote{We assume, for
simplicity, that $B$ is globally defined.  This is true in the NW
model, although it is not the case for compact Lie groups.}.  $S$ may
be viewed as a special case of an action for an open string
propagating on a group manifold and coupled to $A$ at the boundary.

The conditions of conformal invariance of this model in the bulk are
satisfied as in the case of the WZW model on 2-sphere (we ignore the
``back reaction" of the boundary coupling $A$ on the bulk
$\beta$-functions).  The boundary conformal invariance condition
\cite{ACNY},  in general,  may  impose a constraint on $A$ which should
follow from the variation of the Born-Infeld action \cite{ACNY} \
$\int d^{10} x \sqrt{ -\det(G + B + F)_{\mu\nu}} + O(\partial F,
\partial B,R)$.  The leading-order condition of conformal invariance
at the boundary can be represented as the Maxwell equation for $B+F$
on the curved space (group manifold) with metric $G$ 
\begin{equation}
\d_{\mu}\left[{\sqrt G} G^{\mu\nu}G^{\rho\sigma}(F+B)_{\nu\rho}\right]
= 0~.
\end{equation}
In general, the solution for $F$ depends on a specific choice of $B$.
It is easy to check that this condition is indeed satisfied in the NW
model for $B$ as in \eqref{eq:B} and $F_{\mu\nu}=0$.\footnote{It
seems likely that for a ``natural" choice of $B$ this condition is
always satisfied for $F=0$ in a generic WZW model.  This is easy to
check directly for some simple cases like $SU(2)$ and $SL(2,R)$ WZW
models.}

If we vary the  action $S$  we get a bulk term which yields the same
equations of motion as in the $\d\Sigma=0$ case, implying the
conservation of the two sets of currents $J_a$ and $\bar J_a$.  We
also get the  boundary term 
\begin{equation*}
\left. \int_{\d\Sigma} d\tau (g^{-1}\delta g)^a \left[ G_{ab} (g^{-1}
\d_{\sigma} g)^b -i (B_{ab} + F_{ab}) (g^{-1}\d_{\tau}
g)^b\right]\right|^{\sigma=\pi}_{\sigma=0}~,
\end{equation*}
which yields a set of boundary conditions.  The natural question is
whether these conditions are related to the boundary conditions
\eqref{eq:bc}. 

In order to address this question we introduce the coordinates
$x^{\mu}$ on the group manifold (in the particular parametrisation (3)
introduced in Section 2 these are $a_1$, $a_2$, $u$ and $v$), and
consider the left- and right-invariant vielbeins defined by
\begin{equation*}
g^{-1}\d_{\mu}g = {e_{\mu}}^a X_a~, \qquad 
\d_{\mu}g g^{-1} = {\bar e}_{\mu}{}^a X_a~.
\end{equation*}
These vielbeins are related by ${\bar e}_{\mu}{}^a = {e_{\mu}}^b
{C_b}^a$, where $C$ denotes the adjoint action of the group, $g X_a
g^{-1} = {C_a}^b X_b$.

The classical conserved currents are then 
\begin{equation} \label{eq:J(x)}
J_a = -G_{ab}{\bar e}_{\mu}{}^b \d x^{\mu}~, \qquad 
\bar J_a = G_{ab}{e_{\mu}}^b \bar\d x^{\mu}~.
\end{equation}
The surface term in the infinitesimal variation of the action can be
written as
\begin{equation*}
\left.\int_{\d\Sigma}d\tau \delta x^{\mu}
p_{\mu}\right|^{\sigma=\pi}_{\sigma=0}  
\end{equation*}
where $p_{\mu}$ (which is the component of the 2-momentum normal to
the boundary $\d\Sigma$) is given by
\begin{equation*}
p_{\mu} = G_{\mu\nu}\d_{\sigma}x^{\nu} -i(B_{\mu\nu} + F_{\mu\nu})
\d_{\tau}x^{\nu}~, 
\end{equation*}
where $G_{\mu\nu} = {e_{\mu}}^a G_{ab}{e_{\nu}}^b$, $B_{\mu\nu} =
{e_{\mu}}^a B_{ab}{e_{\nu}}^b$ and $F_{\mu\nu} = {e_{\mu}}^a
F_{ab}{e_{\nu}}^b$.  Thus having Neumann boundary conditions in all
directions means imposing $p_{\mu}|_{\d\Sigma}=0$ for all $\mu$.
Using \eqref{eq:J(x)}, we can write $p_{\mu}$ in terms of the
holomorphic and antiholomorphic currents as follows:
\begin{equation*}
p_{\mu} = -[\delta_{\mu}{}^{\rho} - (B+F)_{\mu\nu}G^{\nu\rho}]{\bar
           e}_{\rho}{}^a J_a - [\delta_{\mu}{}^{\rho} +
          (B+F)_{\mu\nu}G^{\nu\rho}]{e_{\rho}}^a \bar J_a~.
\end{equation*}
Then the Neumann boundary conditions take the following compact form
\begin{align}
J + {\mathbf{M}} \bar J &= 0~,\label{eq:cBC}\\
\noalign{where}
{\mathbf{M}} &\equiv \bar e^{-1}
{\frac{\1 + (B+F)G^{-1}}{\1 - (B+F)G^{-1}}}e ~.\label{eq:cBCN}
\end{align}
This is a generalisation of the familiar expression for the boundary
conditions for an open string in a constant $(G,B,F)$ background
\cite{CLNY}.  Note that the matrix  $\mathbf{M}$ is no longer constant
but field-dependent.  Moreover, in the boundary conditions for the Lie
algebra valued currents it is sandwiched between the vielbeins.
  
The field dependent nature of the above matrix allows it to acquire,
for special values of the fields, the $-1$ eigenvalues which are
interpreted as Dirichlet boundary conditions.  In the constant
background (flat) case such a transition from Neumann to Dirichlet
boundary conditions is impossible to attain for any {\it finite}
values of the field $(B+F)G^{-1}$.

Let us now describe the sigma model realisation of a D$p$-brane, with
$p+1$ strictly smaller than the dimension of the target manifold (in
our case $p<3$).  In this case some of the fields satisfy Dirichlet
boundary conditions, and the way we implement this is by imposing
\cite{L} that $x^{\mu}$ at the boundary are given by a set of
functions $\{f^{\mu}\}$ defined on the worldvolume of the D$p$-brane
\begin{equation}
\left. x^{\mu}\right|_{\d\Sigma} = f^{\mu}(y^{m})~,\label{eq:Dbc}
\end{equation}
where $y^m$ denote the coordinates on the $(p+1)$-dimensional
submanifold determined by the worldvolume of the D$p$-brane.  The
infinitesimal variation of the spacetime fields at the boundary is
given by $\left.\delta x^{\mu}\right|_{\d\Sigma} = \d_m f^{\mu}\delta
y^{m}$, so that the boundary term in the variation of the WZW action
becomes
\begin{equation*}
\left.\int_{\d\Sigma}d\tau \delta y^m
p_m\right|^{\sigma=\pi}_{\sigma=0}   ~,
\end{equation*}
\begin{equation*}
p_m = \d_m f^{\mu} G_{\mu\nu}\d_{\sigma}x^{\nu} -i(B +
      F)_{mn}\d_{\tau}y^n~,
\end{equation*}
where $B_{mn} = \d_m f^{\mu} B_{\mu\nu} \d_n f^{\nu}$ and $F_{mn} =
\d_m f^{\mu} F_{\mu\nu} \d_n f^{\nu}$ are the antisymmetric tensor and
the background gauge field induced at  the worldvolume of the
D$p$-brane.  In this case the $p+1$ Neumann boundary conditions become
$p_m|_{\d\Sigma} = 0$ or, in terms of the currents
\begin{equation}\label{eq:NDp}
\d_m f^{\mu}{[\1 - (B+F)G^{-1}]_{\mu}}^{\nu}{\bar e}_{\nu}{}^a J_a +
\d_m f^{\mu}[\1 + (B+F)G^{-1}]_{\mu}{}^{\nu}{e_{\nu}}^a \bar J_a = 0~.
\end{equation}

\subsection{The D3-brane}

Let us compute the matrix of boundary conditions in \eqref{eq:cBC} in
the particular case of the Nappi-Witten background.  The vielbeins $e$
and $\bar e$ are given by
\begin{equation*}
e = \begin{pmatrix}
    \cos u & -\sin u & 0 &  \half a_2\\
    \sin u &  \cos u & 0 & -\half a_1\\
       0   &   0     & 1 &    0      \\
       0   &    0    & 0 &    1
    \end{pmatrix}~,\quad
\bar e = \begin{pmatrix}
         1 &   0  & 0 & -\half a_2\\
         0 &   1  & 0 &  \half a_1\\
       a_2 & -a_1 & 1 & -\half(a_1^2+a_2^2) \\
         0 &    0 & 0 &    1
    	 \end{pmatrix}.
\end{equation*}
If we assume that $F=0$ and we take the antisymmetric tensor field to
be as in \eqref{eq:B} (this choice is consistent with conformal
invariance) we find for  the matrix of the boundary conditions
\begin{equation}\label{eq:solR}
{\mathbf{M}}= \begin{pmatrix}
               \cos u & -\sin u & 0 & 0\\
               \sin u &  \cos u & 0 & 0\\
                  0   &    0    & 1 & 0\\
                  0   &    0    & 0 & 1
              \end{pmatrix}~.\qquad
\end{equation}
Notice that this matrix is of the form \eqref{eq:RI}, with the
parameter $\phi$ given by the field $u$ and $r_1,r_2 = 0$.  Different
choices for $B$ (while keeping $F=0$) will yield different $\mathbf{M}$
(different boundary conditions), but they will still be described by a matrix
of the type $R_I(\phi,r_1,r_2)$ with the parameters given by certain
functions of the fields $a_i,u,v$.

An unusual feature of the boundary conditions described by
\eqref{eq:solR} is their oscillatory character.  In particular, when
$u = (2n+1)\pi$, with $n$=integer, we have two Dirichlet directions,
corresponding to $a_1$ and $a_2$.

It is instructive to rederive these results in a different
parametrisation of the group G
\begin{equation*}
g = e^{x_1 P_1}e^{uJ}e^{x_2 P_1 + vK}~,
\end{equation*}
in which $\{x^\mu\}=\{x_1,x_2,u,v\}$ and the spacetime metric reads 
\begin{equation*}
ds^2 = dx_i dx_i + 2\cos u\ dx_1 dx_2 + b du^2 + 2du dv~.
\end{equation*}
This coordinate system is singular for $u = \pi m$, \ $m$=integer.  If
we choose the antisymmetric tensor to be $B = \half\epsilon_{ij}x_i
dx_j du$ and compute $\mathbf{M}$ we get exactly the same matrix as in
\eqref{eq:solR}.  Here the oscillatory character of the boundary
conditions is very similar to the one of the metric itself.  In
particular, the Dirichlet conditions appear at values of $u$ which
correspond to the singularities of the coordinate system (and the
metric).

\subsection{The D1-brane}

We now consider the sigma model description of the D-string found in
Section 4.  The two Dirichlet boundary conditions are described as in
\eqref{eq:Dbc}, with the functions $f^{\mu}$ given by \eqref{eq:D1}.
Here we are in the static gauge since the two-dimensional worldsheet
of the D-string coincides with the surface defined by the coordinates
$u$ and $v$.

In this case, if we take $F=0$ and $B$ given by \eqref{eq:B}, the
Neumann boundary conditions \eqref{eq:NDp} become
\begin{equation*}
J_3 + \bar J_3 = 0~, \qquad  J_4 + \bar J_4 = 0~,
\end{equation*}
which agree with the Neumann boundary conditions obtained from $R_I$.

It may seem   that we have unnecessarily restricted ourselves to
a particular D-string configuration, namely,  the one given by
\eqref{eq:D1}, instead of considering the most general one in
\eqref{eq:D1a}-\eqref{eq:D1d}.  One can easily derive the Neumann
boundary conditions in the most general case: one obtains a set of two 
relations which agree with the Neumann boundary conditions coming from
the boundary state approach {\it only} for $r_1,r_2=0$.  This result
is not surprising  since
the equations describing the D-string have been obtained under
the assumption that $r_1$ and $r_2$ are constant parameters.  On the
other hand, we have seen in the case of the D3-brane that the two
parameters, $r_1$ and $r_2$, are generically field-dependent.
In order to write down a sigma-model describing a more general
D1-brane configuration one should  probably  consider functions $f^{\mu}$
which are field-dependent.

\subsection{The D0-brane}

Finally, let us describe a sigma model realisation of the D0-brane
configuration obtained in Section 4 and characterised by $R_{II}$.  
In this case the fields $x^{\mu}$ satisfy boundary conditions of the
form \eqref{eq:Dbc}, where the functions $f^{\mu}$ are given by
\eqref{eq:D0a}-\eqref{eq:D0d} (we consider here only the case
$\phi\neq\tfrac{3\pi}2$, the case $\phi=\tfrac{3\pi}2$ being very
similar).  If we compute the Neumann boundary condition \eqref{eq:B},
with $F$ and $B$ as in the previous case, we obtain
\begin{align*}
&(1 + \sin\phi)(J_1 + \bar J_1) + \cos\phi(J_2 + \bar J_2)\\
&\phantom{asdasd} +\tfrac{1}{2}\left[r_1 (1 + \sin\phi) + r_2
\cos\phi\right](J_4 + \bar J_4) = 0~, 
\end{align*}
which is nothing but the Neumann condition corresponding to
$R_{II}$.

\section*{Acknowledgements}

We are grateful to C. Klim\v cik and Y. Lozano for valuable discussions, and
to J.M. Figueroa-O'Farrill for helpful discussions and for drawing the
pictures in the Appendix.  This work was supported in part by PPARC, the
European Commission TMR programme grant ERBFMRX-CT96-0045 and the NSF grant
PHY94-07194.

\appendix

\section*{Appendix A}

Here we describe the D-string configuration in more detail.  It is
convenient to change variables in the Nappi--Witten spacetime to
$x^\mu=(x^0,x^1,x^2,x^3)$ defined in terms of $(a_i)$ as follows:
\begin{align*}
x^0 &= \frac{1}{\sqrt{2}} \left[ a_4 - (1+\half b) a_3 + \tfrac14 (r_1
a_1 + r_2 a_2) \right]\\
x^1 &=a_1\\
x^2 &=a_2\\
x^3 &= \frac{1}{\sqrt{2}} \left[ a_4 + (1-\half b) a_3 \right]~.
\end{align*}
The Nappi--Witten metric in these coordinates becomes
\begin{multline*}
ds^2 = \eta_{\mu\nu} dx^\mu dx^\nu - \tfrac1{32} (r_1 dx^1 +
r_2dx^2)^2 + \tfrac1{2\sqrt{2}} (r_1 dx^1+ r_2 dx^2) dx^0\\
+ \tfrac1{\sqrt{2}} \left(x^1 dx^2 - x^2 dx^1\right) \left(dx^3 - dx^0
+ \tfrac1{4\sqrt{2}} (r_1 dx^1 + r_2 dx^2)\right)~.
\end{multline*}
The virtue of these coordinates is that the intrinsic D-string time
and the time of the ambient spacetime agree.  Indeed,  the explicit
embedding of D-string world sheet   is given by
\begin{align*}
x^0(\sigma,\tau)&=\tau\\
x^1(\sigma,\tau)&=\half r_1 \sin\tfrac1{\sqrt{2}}(\sigma-\tau) + \half
r_2 \left(\cos\tfrac1{\sqrt{2}}(\sigma-\tau) - 1\right)\\
x^2(\sigma,\tau)&= -\half r_1 \left(\cos\tfrac1{\sqrt{2}}(\sigma-\tau)
- 1\right) + \half r_2 \sin\tfrac1{\sqrt{2}}(\sigma-\tau)\\
x^3(\sigma,\tau)&= \sigma - \tfrac1{8\sqrt{2}} \varrho^2
\sin\tfrac1{\sqrt{2}}(\sigma-\tau)~,
\end{align*}
where $\varrho \equiv \sqrt{r_1^2 + r_2^2}$  and 
\begin{align*}
\sigma &= \frac1{\sqrt{2}} \left( \left(1 + \half b + \tfrac18
\varrho^2\right) \alpha + \beta\right)\\
\tau &= \frac1{\sqrt{2}} \left( \left(1 - \half b - \tfrac18
\varrho^2\right) \alpha - \beta\right)~,
\end{align*}
relative to which the metric \eqref{eq:Dmet} becomes simply $ds^2 =
d\sigma^2 - d\tau^2~$. 

Therefore the $x^0=\mathrm{constant}$ hyperplane cuts the
D-string worldsheet in the D-string itself.  The following snapshots
illustrate the D-string embedded in the three-dimensional hypersurface
$x^0 = 0$, for different values of $\varrho$.

\begin{figure}[h!]
\centering
\begin{tabular}{ccccccc}
\epsfig{file=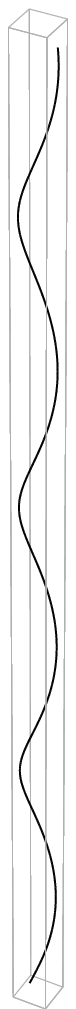,height=6cm} &
\epsfig{file=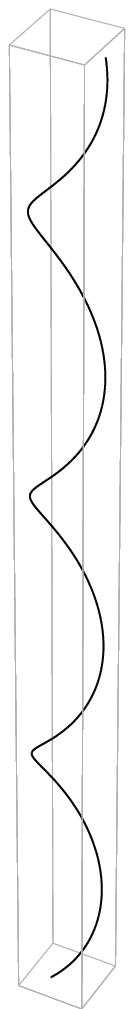,height=6cm} & 
\epsfig{file=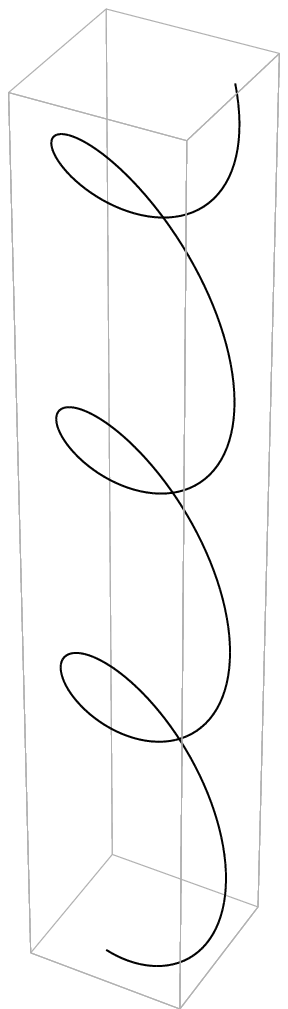,height=6cm} &
\epsfig{file=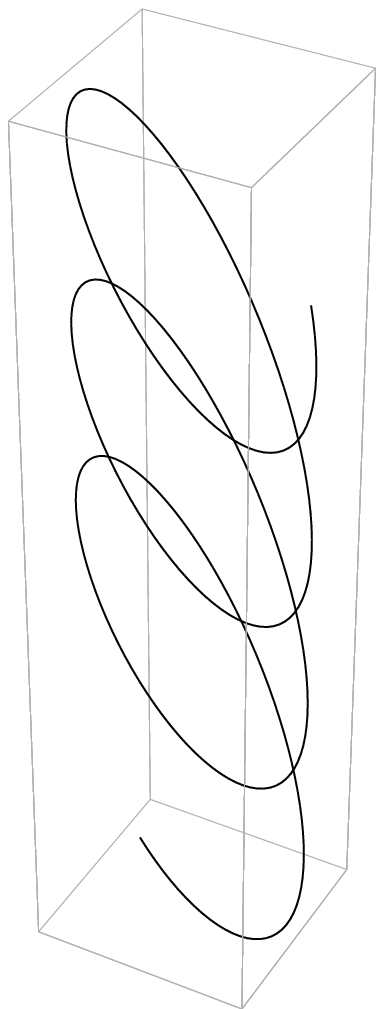,height=6cm} & 
\epsfig{file=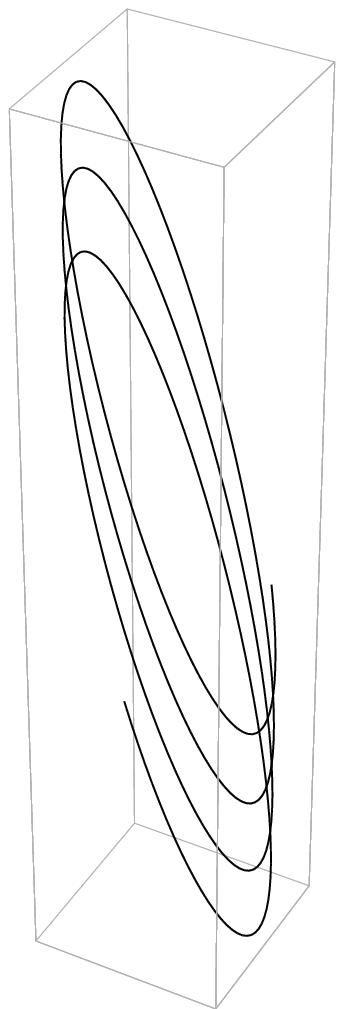,height=6cm} & 
\epsfig{file=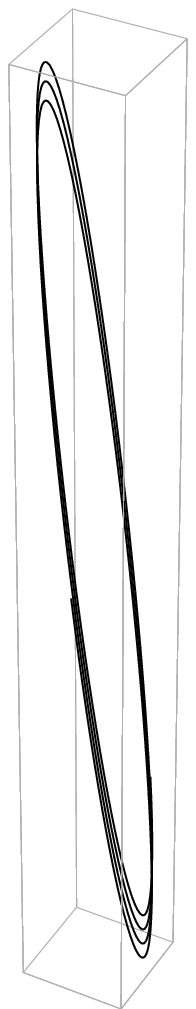,height=6cm} &
\epsfig{file=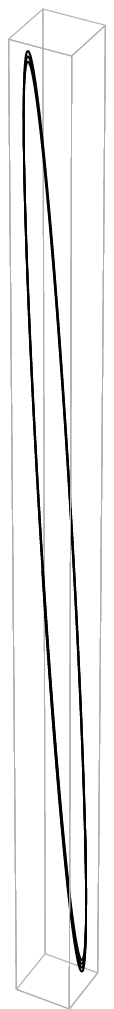,height=6cm}
\end{tabular}
\caption{$x^0=\tau=0$ snapshots for $\varrho=1,2,5,10,20,50,100$.}
\end{figure}

\section*{Appendix B}

The above discussion of the bosonic NW  model admits an $N{=}2$
generalisation.  Indeed,  the $N{=}1$ extension of \eqref{eq:wzw}
describes a superconformal theory with the superconformal algebra
generated by $(\sT,\sG)$.  Further, if we consider the algebra
\eqref{eq:gg} and we define $t_+ = \{P_+^1, P_+^2\}$ and $t_- =
\{P_-^1, P_-^2\}$, where 
\begin{equation}
P_{\pm}^1 = \half (P_1 \mp iP_2)~, \qquad
P_{\pm}^2 = \half (J \mp i K)~,
\end{equation}
we can easily check that $[t_{\pm},t_{\pm}]\subset t_{\pm}$, 
 which is
the condition for the $N{=}1$ theory to admit an $N{=}2$ extension,
with the superconformal algebra generated by $(\sT,\sG^{\pm},\sJ)$.
In this case we can apply the boundary state approach as described in
\cite{SDKS} and determine which of the solutions that we have 
found for the bosonic boundary states ($R_I$, $R_{II}$) survive as
boundary states of the $N{=}2$ model.  The boundary conditions in the
$N{=}2$ case fall into the two types, A and B, and the additional
requirement that the boundary state described by $R$ has to satisfy
can be stated as follows:

\begin{itemize}
\item[(i)] A--type boundary conditions
\begin{equation}
R\ A = -A\ R~;
\end{equation}
\item[(ii)]B--type boundary conditions
\begin{equation}
R\ A = A\ R~,
\end{equation}
\end{itemize}
where $A$ is the complex structure on $\gg$, which in the NW  case can
be taken to be 
\begin{equation}
A = \begin{pmatrix}
	 0 & 1 &  0 & 0\\
        -1 & 0 &  0 & 0\\
         0 & 0 &  0 & 1\\
         0 & 0 & -1 & 0
	 \end{pmatrix}~.
\end{equation}
We can now check explicitly the existence of each of the solutions ($R_I$,
$R_{II}$) found in Section 3.  We find that $R_{II}$ yields no solutions for
either of the two types of boundary conditions, whereas $R_I$ only yields
solutions for the B-type boundary conditions. These are given by
\eqref{eq:RI} with $r_1,r_2=0$, i.e.  are of the form $R_I(\phi,0,0)$.

%
\providecommand{\href}[2]{#2}\begingroup\raggedright\endgroup
\end{document}